# Tuning the spin Hall effect of Pt from the moderately dirty to the superclean regime

Edurne Sagasta[1,*], Yasutomo Omori[2,*], Miren Isasa[1,*], Martin Gradhand[3], Luis E. Hueso[1,4], Yasuhiro Niimi[2,5], YoshiChika Otani[2,6], Fèlix Casanova[1,4,†]

[*]These authors contributed equally to this work
[†]Corresponding author: f.casanova@nanogune.eu

[1]CIC nanoGUNE, 20018 Donostia-San Sebastian, Basque Country, Spain
[2]Institute for Solid State Physics, University of Tokyo, Kashiwa, Chiba 277-8581, Japan
[3]H. H. Wills Physics Laboratory, University of Bristol, Bristol BS8 1TL, United Kingdom
[4]IKERBASQUE, Basque Foundation for Science, 48011 Bilbao, Basque Country, Spain
[5]Department of Physics, Graduate School of Science, Osaka University, 1-1 Machikaneyama, Toyonaka, Osaka 560-0043, Japan
[6]RIKEN-CEMS, 2-1 Hirosawa, Wako, Saitama 351-0198, Japan

## Abstract

We systematically measure and analyze the spin diffusion length and the spin Hall effect in Pt with a wide range of conductivities using the spin absorption method in lateral spin valve devices. We observe a linear relation between the spin diffusion length and the conductivity, evidencing that the spin relaxation in Pt is governed by the Elliott-Yafet mechanism. We find a single intrinsic spin Hall conductivity ($\sigma_{SH}^{int}$=1600±150 $\Omega^{-1}$cm$^{-1}$) for Pt in the full range studied which is in good agreement with theory. For the first time we have obtained the crossover between the moderately dirty and the superclean scaling regimes of the spin Hall effect by tuning the conductivity. This is equivalent to that obtained for the anomalous Hall effect. Our results explain the spread of the spin Hall angle values in the literature and find a route to maximize this important parameter.

Spin-orbit interaction is an essential ingredient in solid state physics [1,2] that has been gaining interest in the last decade due to the advantages it offers to exploit the coupling between spin and orbital momentum of electrons in *spintronic* devices, leading to the emerging field of *spin-orbitronics* [3]. The discovery of new charge-to-spin current conversion effects such as the spin Hall effect (SHE) [4,5,6,7], the Rashba-Edelstein effect (REE) [8,9,10] or the spin-momentum locking (SML) in topological insulators [11,12,13] is expanding the possibility to create and detect spin currents without using ferromagnets (FM) or magnetic fields. For instance, magnetization switching of ferromagnetic elements has been recently achieved with torques arising from SHE [14], REE [15] or SML [16], and new spin-dependent phenomena such as the spin Seebeck effect [17] or spin pumping [18] have been discovered by using SHE to detect spin currents.

The SHE is thus the crucial effect behind this breakthrough. Although it was predicted theoretically by Dyakonov and Perel 45 years ago [1] and revisited by Hirsch in 1999 [4], it took a bit longer to observe the first direct experimental evidences in semiconductors [19] and metals [6,7,18]. The SHE in a non-magnet (NM) basically shares the same origin as the anomalous Hall effect (AHE) in FMs: in both effects, the spin-orbit coupling generates the opposite deflection of the spin-up and spin-down electrons in a charge current, leading to a transverse spin current. This can be detected as a transverse voltage in the AHE because the intrinsic spin polarization in FMs gives rise to a net charge accumulation. On the other hand, the SHE creates a transverse pure spin current, making it more difficult to detect. The inverse spin Hall effect (ISHE) is the reciprocal effect, in which pure spin currents are converted into charge currents. The efficiency of these conversions is given by the spin Hall angle, $\theta_{SH}$.

The mechanisms behind the SHE, which can be either intrinsic or extrinsic, were first studied in the framework of the AHE. The intrinsic contribution, first proposed by Karplus and Luttinger [20], relies on the spin-dependent band structure of the conductor and the transverse displacement of the spin-up and spin-down electrons occurs in between scattering events. Skew scattering, proposed by Smit [21], is an extrinsic contribution where the spin-dependent scattering arises due to the effective spin-orbit coupling of impurities in the lattice. A phonon skew scattering [22,23] has recently been shown to contribute in metals such as Au [24]. Side jump, another extrinsic mechanism introduced by Berger [25], results in a spin-dependent sideway displacement upon repeated scattering.

In contrast to the AHE [22,26,27,28], a systematic experimental study of the different mechanisms contributing to the SHE for relevant materials is lacking. Finding routes to maximize the SHE is not possible as long as it remains unclear whether the dominant mechanism in a material is intrinsic or extrinsic. This issue has particularly been controversial in Pt, the prototypical SHE metal, although a consensus that the intrinsic contribution is the dominant one is emerging [24,29,30]. Nevertheless, there is still a significant spread in the $\theta_{SH}$ values of Pt among different groups and techniques [31,32]. The spin-memory loss explained by Rojas-Sánchez *et al.* [31] is one of the causes for this, as well as the wide range of spin diffusion lengths $\lambda_{Pt}$ used by different groups [32]. A proper understanding of the SHE in order to correctly quantify and tune $\theta_{SH}$ in Pt is thus of utmost importance.

In this Letter, we experimentally study the spin diffusion length and the SHE in Pt with a broad range of longitudinal conductivities ($\sigma_{Pt}$). The linear relation obtained between $\lambda_{Pt}$ and $\sigma_{Pt}$ evidences that the spin relaxation in Pt is governed by the Elliott-Yafet mechanism. We obtain a single intrinsic spin Hall conductivity for the full range studied, in good agreement with theoretical predictions. In addition, we identify slightly different skew scattering angles for Pt deposited with different techniques. By tuning the SHE with $\sigma_{Pt}$, we can thus observe the crossover from a moderately dirty regime, where the intrinsic contribution is dominant, to the superclean regime, dominated by extrinsic effects. A similar crossover has been discussed in the AHE [27,28], but has never been observed experimentally in any spin Hall system before. This result clearly elucidates the detailed mechanism of the SHE in its prototypical metal.

We used the spin absorption method in lateral spin valve devices, which enables us to quantitatively derive both the spin diffusion length (via the spin absorption) and the spin Hall angle (via the ISHE) of Pt on the same device [7,24,29,33,34,35,36]. To this end, eight devices were fabricated on top of a $SiO_2$(150nm)/Si substrate by using multiple-step e-beam lithography, subsequent metal deposition and lift-off. Each device contains two Py/Cu lateral spin valves, both with the same Py interelectrode distance $L \sim 630$nm, but one of them with a Pt wire in between the electrodes, as shown in Fig. 1(a). First, each pair of Py electrodes were patterned with different widths, ~100nm and ~170nm, in order to obtain different switching magnetic fields, and 35nm of Py were e-beam evaporated. During the second step, a ~130-nm-wide and 20-nm-thick Pt wire was deposited by e-beam evaporation (base pressure $\leq 1 \times 10^{-8}$torr, rate 0.1-2Å/s, substrate temperature 5-7ºC) in half of the devices (E1,E2,E3,E4) and by magnetron sputtering (base pressure $1\times10^{-7}$-$1\times10^{-8}$torr, power 80 W, Ar pressure $3\times10^{-3}$torr, rate 1.3Å/s, substrate temperature 25ºC) in the other half (S1,S2,S3,S4). The different Pt wires cover a broad range of resistivities, with evaporated ones having a smaller residual resistivity $\rho_{Pt,0}$ than sputtered ones, see Table I [37]. In the third lithography step, a ~150-nm-wide channel was patterned and a 100-nm-thick Cu was thermally evaporated. In order to have highly transparent Py/Cu and Pt/Cu interfaces, the surfaces of the Py and Pt wires were cleaned via Ar-ion beam etching before the Cu deposition. All non-local transport measurements described below were carried out in a liquid-He cryostat (applying an external magnetic field $H$ and varying temperature $T$) using a "dc reversal" technique [39].

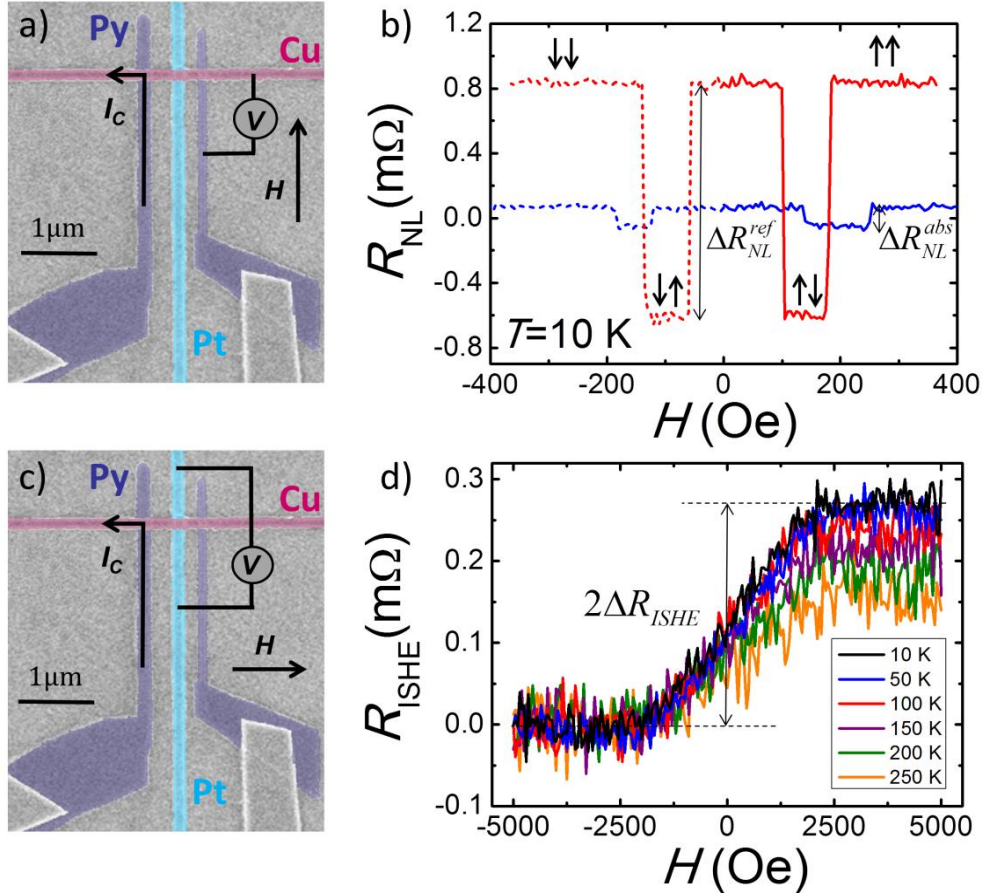

FIG. 1. a) SEM image of a Py/Cu lateral spin valve with a Pt wire between the two Py electrodes. The non-local measurement configuration, the direction of the applied magnetic field ($H$) and the materials (Py, Cu and Pt) are shown. b) Non-local resistance as a function of $H$ measured at $I_C$ =100µA and 10K in the configuration shown in a) for a Py/Cu lateral spin valve with (blue line) and without (red line) a Pt wire in between the Py electrodes. The solid (dashed) line corresponds to the increasing (decreasing) magnetic field. The reference spin signal ($\Delta R_{NL}^{ref}$) and the spin signal with Pt absorption ($\Delta R_{NL}^{abs}$) are tagged. c) SEM image of the same device used now to measure the ISHE. The materials, the direction of $H$ and the measurement configuration for ISHE are shown. d) ISHE resistance as a function of $H$ measured at $I_C$ =100µA and different temperatures in the configuration shown in c). The ISHE signal ($2\Delta R_{ISHE}$) for 10K is tagged. Images and data correspond to device S2.

When a spin-polarized current is injected from one Py electrode, as shown in Fig. 1(a), a spin accumulation is created at the Py/Cu interface. This spin accumulation diffuses along both sides of the Cu channel, creating a pure spin current which is detected as a voltage by the second Py electrode. Normalizing the measured voltage to the injected current $I_C$, the non-local resistance $R_{NL}$ is defined. This value changes sign when the relative magnetization of the two Py electrodes is switched from parallel to antiparallel by sweeping $H$. The change from positive to negative $R_{NL}$ is defined as the spin signal $\Delta R_{NL}$. The reference value $\Delta R_{NL}^{ref}$ is measured without the Pt wire (see red line in Fig. 1(b)). If a Pt wire is inserted between the Py electrodes, a part of the spin current diffusing along the Cu channel will be absorbed into the Pt and a smaller spin signal $\Delta R_{NL}^{abs}$ will be measured in the Py detector, as shown by the blue line in Fig. 1(b). The spin diffusion length of Pt is obtained from the ratio of both spin signals, which from the one-dimensional spin diffusion model for transparent interfaces is expressed as:

$$\frac{\Delta R_{NL}^{abs}}{\Delta R_{NL}^{ref}} = \frac{2Q_{Pt}\left[\sinh\left(\frac{L}{\lambda_{Cu}}\right)+2Q_{Py}e^{\left(\frac{L}{\lambda_{Cu}}\right)}+2Q_{Py}^2 e^{\left(\frac{L}{\lambda_{Cu}}\right)}\right]}{\cosh\left(\frac{L}{\lambda_{Cu}}\right)-\cosh\left[\frac{L-2d}{\lambda_{Cu}}\right]+2Q_{Py}\sinh\left[\frac{d}{\lambda_{Cu}}\right]e^{\frac{L-d}{\lambda_{Cu}}}+2Q_{Pt}\sinh\left[\frac{L}{\lambda_{Cu}}\right]+4Q_{Py}Q_{Pt}e^{\frac{L}{\lambda_{Cu}}}+2Q_{Py}e^{\frac{d}{\lambda_{Cu}}}\sinh\left[\frac{(L-d)}{\lambda_{Cu}}\right]+2Q_{Py}^2 e^{\frac{L}{\lambda_{Cu}}}+4Q_{Py}^2 Q_{Pt}e^{\frac{L}{\lambda_{Cu}}}} \quad (1)$$

where $Q_{Py(Pt)} = \frac{R_{Py(Pt)}}{R_{Cu}}$, being $R_{Cu} = \frac{\lambda_{Cu}\rho_{Cu}}{w_{Cu}t_{Cu}}$, $R_{Py} = \frac{\lambda_{Py}\rho_{Py}}{w_{Cu}w_{Py}(1-\alpha_{Py}^2)}$ and $R_{Pt} = \frac{\lambda_{Pt}\rho_{Pt}}{w_{Cu}w_{Pt}\tanh\frac{t_{Pt}}{\lambda_{Pt}}}$ the spin resistances of the Cu channel, Py electrodes and Pt wire, respectively. $\rho_{Cu,Py,Pt}$, $\lambda_{Cu,Py,Pt}$, $w_{Cu,Py,Pt}$, and $t_{Cu,Pt}$ are the resistivites, spin diffusion lengths, widths and thicknesses, respectively. The Pt resistivities for all devices are plotted as a function of temperature in Fig. 2(a). $\alpha_{Py}$ is the spin polarization of Py. $L$ is the distance between the two Py electrodes, whereas $d$ is the distance between the Py injector and the Pt wire. Since $R_{Cu}$ and $R_{Py}$ values are well known from our previous work [40,41], $\lambda_{Pt}$ can be obtained from Eq. (1) [42]. By repeating the measurement shown in Fig. 1(b) at different temperatures and devices, $\lambda_{Pt}$ as a function of temperature was obtained for all devices (Fig. 2(b)).

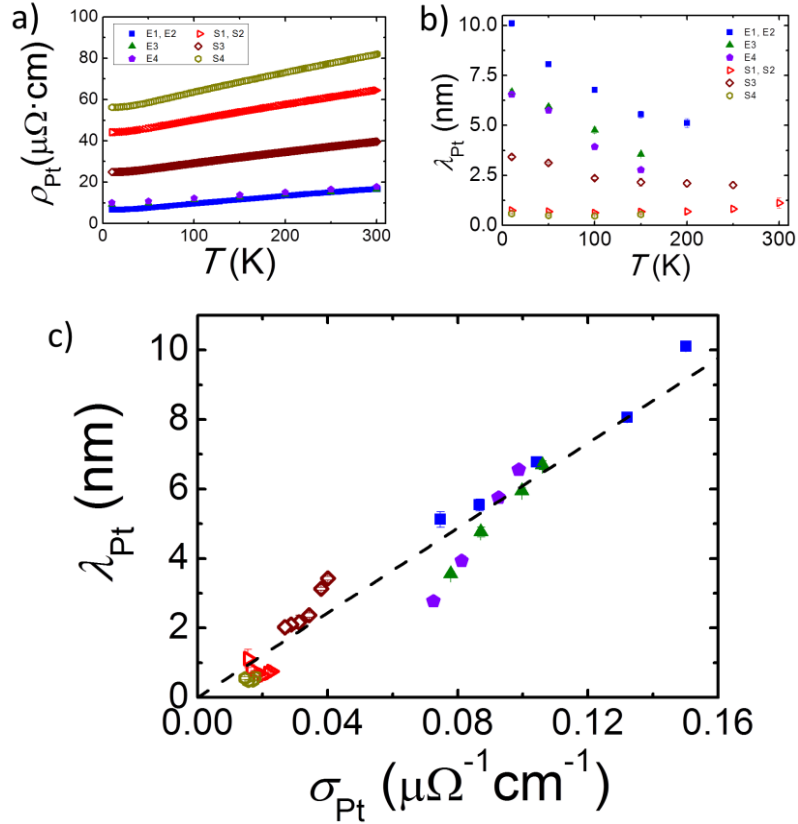

FIG. 2. a) Resistivity and b) spin diffusion length of Pt as function of temperature for all devices. Error bars are included. c) Spin diffusion length of Pt as a function of longitudinal conductivity for all devices. Error bars are included. Black dashed line is a linear fitting of the experimental data. Since devices E1 and E2 (S1 and S2)

were fabricated in the same chip, Pt was evaporated (sputtered) in the same deposition, hence it is assumed that E1 and E2 (S1 and S2) have the same $\rho_{Pt}$ and $\lambda_{Pt}$.

The spin diffusion lengths are plotted in Fig. 2(c) as a function of the conductivity of Pt ($\sigma_{Pt} = \rho_{Pt}^{-1}$), which changes from device to device and with temperature. The linear dependence between $\lambda_{Pt}$ and $\sigma_{Pt}$ confirms that Elliott-Yafet [46,47] is the main spin relaxation mechanism in Pt, as also observed recently using other experimental techniques [30,48]. From our data, we obtain a slope of $(0.61\pm0.02)\times10^{-15}\Omega m^2$, which is in excellent agreement with a recent theoretical prediction (($0.63\pm0.02)\times10^{-15}\Omega m^2$) from first-principles scattering theory combined with temperature-induced disorder [49]. The spin-flip probability for Pt can be calculated from the slope [50], yielding $a_{sf}$=0.57, a large value expected from the strong spin-orbit coupling in Pt [52]. For comparison, a good spin transport metal such as Cu has $a_{sf}$=4.9-11.0×10$^{-4}$ [40,53].

Next, we measured the ISHE in Pt for the eight devices by changing the measurement configuration to the one described in Fig. 1(c). As in the previous configuration, the pure spin current injected from the Py electrode diffuses along the Cu channel and is partly absorbed by the Pt wire. In this Pt wire, due to the ISHE, a charge current perpendicular to both the spin current direction and spin polarization is created and, thus, a voltage drop is generated along the Pt wire. The measured voltage normalized to the injected current $I_C$ yields the ISHE resistance $R_{ISHE}$. By switching the orientation of the magnetic field, the opposite $R_{ISHE}$ is obtained, since the Py magnetization is inverted as well as the orientation of the spin polarization. The difference of the two $R_{ISHE}$ values is twice the ISHE signal: $2\Delta R_{ISHE}$. Figure 1(d) shows the measured $R_{ISHE}$ at different temperatures for one device. We repeated these measurements for each device.

$\Delta R_{ISHE}$ is related to the spin Hall conductivity $\sigma_{SH}$ by [29]:

$$\sigma_{SH} = \sigma_{Pt}^2 \frac{w_{Pt}}{x_{Pt}} \left(\frac{I_C}{\overline{I_s}}\right) \Delta R_{ISHE} \quad (2)$$

where $x_{Pt}$ is the shunting factor which takes into account the current in the Pt that is shunted through the Cu and is obtained from numerical calculations using a finite elements method [54]. $\overline{I_s}$ is the effective spin current that contributes to the ISHE in Pt and is given by [24]:

$$\frac{\overline{I_s}}{I_c} = \frac{\lambda_{Pt}\left(1-e^{-\frac{t_{Pt}}{\lambda_{Pt}}}\right)^2}{t_{Pt}\left(1-e^{-\frac{2t_{Pt}}{\lambda_{Pt}}}\right)} \cdot \frac{2\alpha_{Py}[Q_{Py}\sinh[(L-d)/\lambda_{Cu}]+Q_{Py}^2 e^{(L-d)/\lambda_{Cu}}]}{\cosh\left(\frac{L}{\lambda_{Cu}}\right)-\cosh\left[\frac{L-2d}{\lambda_{Cu}}\right]+2Q_{Py}\sinh\left[\frac{d}{\lambda_{Cu}}\right]e^{\frac{L-d}{\lambda_{Cu}}}+2Q_{Pt}\sinh\left[\frac{L}{\lambda_{Cu}}\right]+4Q_{Py}Q_{Pt}e^{\frac{L}{\lambda_{Cu}}}+2Q_{Py}e^{\frac{d}{\lambda_{Cu}}}\sinh\left[\frac{(L-d)}{\lambda_{Cu}}\right]+2Q_{Py}^2 e^{\frac{L}{\lambda_{Cu}}}+4Q_{Py}^2 Q_{Pt}e^{\frac{L}{\lambda_{Cu}}}} \quad (3)$$

The spin Hall resistivity $\rho_{SH}$ is related to $\sigma_{SH}$ as $\rho_{SH} = -\sigma_{SH}/(\sigma_{Pt}^2 + \sigma_{SH}^2) \approx -\sigma_{SH}/\sigma_{Pt}^2$. The spin Hall angle $\theta_{SH}$ can be written in terms of either $\sigma_{SH}$ or $\rho_{SH}$: $\theta_{SH} = \sigma_{SH}/\sigma_{Pt} = -\rho_{SH}/\rho_{Pt}$. In order to analyze how the intrinsic and extrinsic mechanisms contribute to the SHE in each device, we will analyze the dependence of $\rho_{SH}$ on the longitudinal resistivity $\rho_{Pt}$. As the side jump mechanism arises only in materials with high impurity concentrations [5,55,56], this contribution is negligible in our high purity Pt. The same holds for the phonon skew scattering, due to its small effect in Pt [24]. Therefore, based on the scaling relation introduced by Tian *et al.* [22], this leads us to:

$$-\rho_{SH} = \alpha_{ss}\rho_{Pt,0} + \sigma_{SH}^{int}\rho_{Pt}^2 \quad (4)$$

By plotting $-\rho_{SH}$ against $\rho_{Pt}^2$, we are able to fit a linear function with a slope that corresponds to the intrinsic contribution, $\sigma_{SH}^{int}$, and the intercept divided by $\rho_{Pt,0}$ defines the skew scattering angle $\alpha_{ss}$. Figure 3 shows the data for all devices and the corresponding linear fits. The values extracted from the eight devices are collected in Table I.

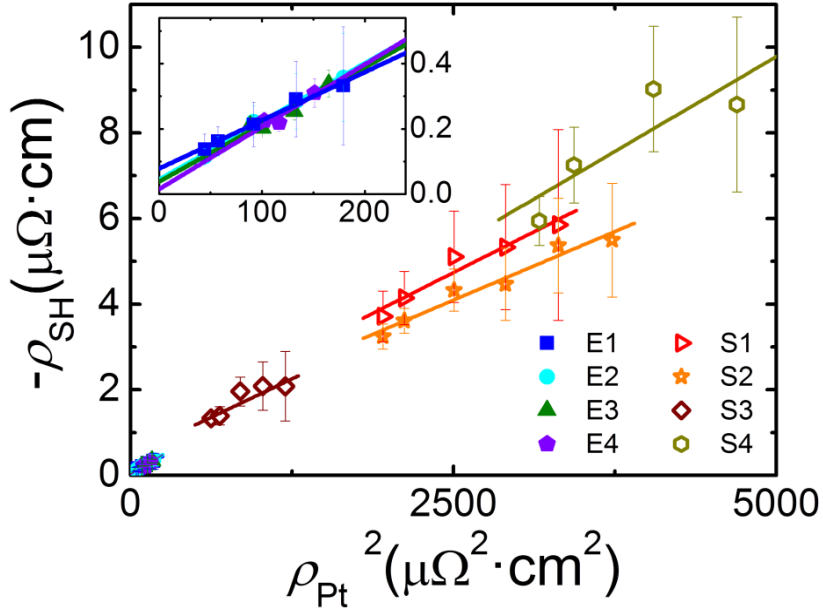

FIG. 3. Spin Hall resistivity as a function of the square of the longitudinal resistivity for all devices. Error bars are included. Solid lines correspond to the fit of the data to Eq. (4). Inset: Zoom of the previous plot showing the data of the devices with evaporated Pt.

Interestingly, the data in Table I reveals that the extracted intrinsic spin Hall conductivities for all the devices are very close to each other, especially when taking into account the different resistivities and $\theta_{SH}$ in each device. We obtain an average value of $\sigma_{SH}^{int}$=1600±150$\Omega^{-1}cm^{-1}$ for Pt, indicating that the intrinsic contribution of the spin Hall conductivity is a constant within a 10% dispersion. This is a remarkable finding, which is in excellent agreement with theoretical values of 1300$\Omega^{-1}cm^{-1}$ [57] and 1600$\Omega^{-1}cm^{-1}$ [58] obtained with different approaches. The predicted decrease of the intrinsic spin Hall conductivity of Pt at higher resistivities by Tanaka et al. [57] lies outside our studied range. A recent experimental study employing the spin torque ferromagnetic resonance technique reports a lower bound of $\sigma_{SH}^{int}$=2950±100$\Omega^{-1}cm^{-1}$ for Pt [30], much higher than ours and the theoretical predictions.

The skew scattering angle yields similar values for all the devices deposited with the same technique, but slightly different for each deposition type. The observation is reasonable as this extrinsic contribution depends directly on the kind of defects in the Pt. Sputtered and evaporated Pt have different grain sizes and, moreover, the deposition in different chambers gives rise to the presence of different impurities, hence explaining the different skew scattering contribution in each type of Pt.

Table I. Intrinsic spin Hall conductivity ($\sigma_{SH}^{int}$) and skew scattering angle ($\alpha_{ss}$) extracted from the individual fittings of each device used in this work. Residual resistivity ($\rho_{Pt,0}$), the spin diffusion length ($\lambda_{Pt}$) and the spin Hall angle ($\theta_{SH}$) at 10K are also included.

| device | $\rho_{Pt,0}(\mu\Omega cm)$ | $\lambda_{Pt}(nm)$ | $\theta_{SH}(\%)$ | $\sigma_{SH}^{int}(\Omega^{-1}cm^{-1})$ | $\alpha_{ss}(\%)$ |
|---|---|---|---|---|---|
| E1 | 6.66 | 10.1±0.1 | 2.1±0.7 | 1480±110 | 1.2±0.2 |
| E2 | 6.66 | 10.1±0.1 | 1.7±0.4 | 1780±95 | 0.7±0.2 |
| E3 | 9.42 | 6.7±0.1 | 2.2±0.2 | 1750±360 | 0.4±0.5 |
| E4 | 10.12 | 6.5±0.1 | 2.2±0.3 | 1910±700 | 0.1±0.9 |
| S1 | 44.19 | 0.75±0.03 | 8.5±1.3 | 1525±220 | 2.1+1.3 |
| S2 | 44.19 | 0.75±0.03 | 7.4±0.7 | 1280±140 | 2.0±0.9 |
| S3 | 24.96 | 3.43±0.05 | 5.3±0.6 | 1435±390 | 1.9±1.3 |
| S4 | 56.25 | 0.59±0.01 | 10.7±1.0 | 1770±760 | 1.6±5.2 |

In contrast to Ref. 22, we cannot plot a universal curve for all devices using Eq. (4) because the extrinsic contribution differs from the evaporated to the sputtered Pt. Nevertheless, we can still plot $\theta_{SH}$ (and $\sigma_{SH}$) as a function of $\sigma_{Pt}$ (see Fig. 4 and inset) in order to compare the relative weight of the different contributions in an analogy to the different scaling regimes observed in the AHE [22,26,27,28].

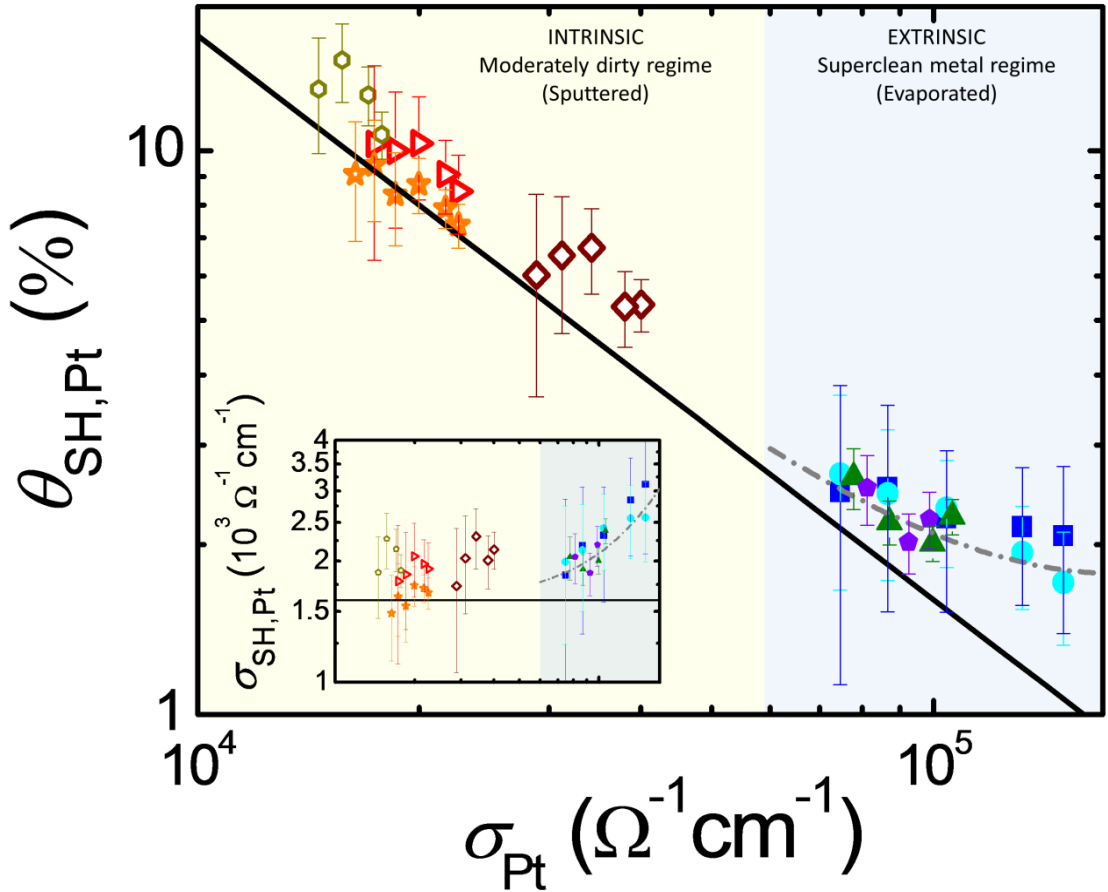

FIG. 4. Spin Hall angle as a function of the longitudinal conductivity of Pt for all devices. Error bars are included. The regions with different scaling regimes are indicated. Black solid line corresponds to the intrinsic contribution of the spin Hall angle $\theta_{SH}^{int} = \sigma_{SH}^{int}/\sigma_{Pt}$, using $\sigma_{SH}^{int}=1600\Omega^{-1}$cm$^{-1}$. Grey dashed line in the superclean region corresponds to the total spin Hall angle calculated with both intrinsic and skew scattering contributions, using the average value $\alpha_{ss}(\%)=0.6$ obtained for this region. Inset: Same data plotted as spin Hall conductivity. The scale of the horizontal axis is the same as in the main panel.

The spin Hall angle for evaporated and sputtered devices scale in a very different way with $\sigma_{Pt}$, as can be seen in Fig. 4. $\theta_{SH}$ for sputtered devices, with highest resistivity, shows the same trend expected from the intrinsic contribution ($\theta_{SH}^{int} = \sigma_{SH}^{int}/\sigma_{Pt}$, black solid line), and the total experimental $\theta_{SH}$ nearly merges into the intrinsic value (the small difference is given by the minor contribution of the skew scattering). This region dominated by the intrinsic scaling regime thus corresponds to the moderately dirty region, similarly to what is observed in the AHE [27,28]. In contrast, in the lower resistivity region, the intrinsic contribution cannot explain the values of the experimental data, even the trend. Nevertheless, by adding the corresponding extrinsic contribution for this region to the diminishing intrinsic one, we obtain the grey dashed line that matches perfectly with our data. This region is thus representing a clean metal, where the skew scattering dominates the scaling. Consequently, for the first time we observe the crossover from the intrinsic moderately dirty regime to the extrinsic superclean regime for the SHE, demonstrating a perfect correspondence with the AHE [27,28].

To conclude, we experimentally show for Pt a general scaling of the SHE. We demonstrate that $\sigma_{SH}^{int}$ is constant in Pt and this allows us to move from an intrinsic to an extrinsic regime when decreasing the resistivity from a moderately dirty to a clean metal. It is a further step towards a complete understanding of the SHE phenomenon, which can be extrapolated to other materials with strong spin-orbit coupling showing SHE. Interestingly, our experimental results evidence that the variation of the Pt resistivity among different groups is one of the main reasons of the spread of $\theta_{SH}$ values in literature. Indeed, we are able to tune $\theta_{SH}$ from ~2 to 14% by varying the Pt resistivity from ~7 to 70μΩcm. A very important consequence is that we show a clear path to enhance $\theta_{SH}$ by simply increasing the resistivity of any material with a dominant intrinsic contribution to the SHE. Additionally, we confirmed that Elliott-Yafet is the main spin relaxation mechanism in Pt.

ACKNOWLEDGMENTS

The authors thank Dr. P. Laczkowski for fruitful discussions. This work is supported by the European Research Council (257654-SPINTROS), by the Spanish MINECO under Projects No. MAT2012-37638 and MAT2015-65159-R and by the Japanese Grant-in-Aid for Scientific Research on Innovative Area, "Nano Spin Conversion Science" (Grant No. 26103002). M.G. acknowledges financial support from the Leverhulme Trust via an Early Career Research Fellowship (ECF-2013-538). E.S and M.I thank the Spanish Ministry of Education, Culture and Sport and the Basque Government, respectively, for a Ph.D. fellowship (Grants No. FPU14/03102 and No. BFI-2011.106). Y. Omori acknowledges financial support from JSPS through "Research program for Young Scientists" and "Program for Leading Graduate Schools (MERIT)".